\documentclass[%
aps,prb,superscriptaddress,twocolumn,floatfix,10pt,longbibliography]{revtex4-2}
\usepackage[utf8]{inputenc}
\usepackage{amsmath,amssymb}
\usepackage{empheq}
\usepackage{float}
\usepackage{lipsum}
\usepackage{mathtools,cuted}
\usepackage{esvect}
\usepackage{multirow}
\usepackage{xcolor}
\usepackage{soul}
\usepackage[export]{adjustbox}
\usepackage{graphicx}
\usepackage{dcolumn}
\usepackage{bm}
\usepackage{hyperref}
\usepackage{physics}
\usepackage[mathlines]{lineno}
\usepackage[capitalise]{cleveref}
\usepackage{bbm}
\usepackage{orcidlink}
\usepackage{babel}
\usepackage{dsfont}
\usepackage{booktabs}
\usepackage{pifont}

\makeatletter
\def\bbl@set@language#1{%
	\edef\languagename{%
		\ifnum\escapechar=\expandafter`\string#1\@empty
		\else\string#1\@empty\fi}%
	\@ifundefined{babel@language@alias@\languagename}{}{%
		\edef\languagename{\@nameuse{babel@language@alias@\languagename}}%
	}%
	\select@language{\languagename}%
	\expandafter\ifx\csname date\languagename\endcsname\relax\else
	\if@filesw
	\protected@write\@auxout{}{\string\select@language{\languagename}}%
	\bbl@for\bbl@tempa\BabelContentsFiles{%
		\addtocontents{\bbl@tempa}{\xstring\select@language{\languagename}}}%
	\bbl@usehooks{write}{}%
	\fi
	\fi}
\newcommand{\DeclareLanguageAlias}[2]{%
	\global\@namedef{babel@language@alias@#1}{#2}%
}
\makeatother

\DeclareLanguageAlias{en}{english}
\DeclareLanguageAlias{English}{english}
\DeclareLanguageAlias{Englisch}{english}
\DeclareLanguageAlias{EN}{english}
\DeclareLanguageAlias{en-US}{english}

\begin{document}
	
\title{Topological states in fractals through a self-similar tensor network encoding}
	
\author{Zebedeus F. Osseweijer\,\orcidlink{0009-0004-0537-7592}}
\affiliation{Institute of Theoretical Physics, Utrecht University, Utrecht, 3584 CC, Netherlands}
\affiliation{Department of Applied Physics, Aalto University, 02150 Espoo, Finland}

\author{Jagoda Malag}
\affiliation{Department of Applied Physics, Aalto University, 02150 Espoo, Finland}
\affiliation{Wroc{\l}aw University of Science and Technology, 50-370 Wroc{\l}aw, Poland}

\author{Jose L. Lado\,\orcidlink{0000-0002-9916-1589}}
\affiliation{Department of Applied Physics, Aalto University, 02150 Espoo, Finland}

\author{Tiago V. C. Ant\~ao\,\orcidlink{0000-0003-3622-2513}}
\affiliation{Department of Applied Physics, Aalto University, 02150 Espoo, Finland}

\author{Anouar Moustaj\,\orcidlink{0000-0002-9844-2987}}
\affiliation{Department of Applied Physics, Aalto University, 02150 Espoo, Finland}

\date{\today}
	
\begin{abstract}
Tensor network methods have been instrumental in addressing exponentially large many-body problems. Most recently, these techniques have also been applied to single-particle tight-binding Hamiltonians, allowing observables to be evaluated on exponentially large lattices.
Here, we show that the scale-separated nature of these methods allows for a natural encoding of self-similar structures, such as the Sierpi\'nski gasket, a scale-invariant fractal with Hausdorff dimension $\log 3/\log 2\approx1.58$, yielding an exact and compact representation of Hamiltonians on such lattices. Because the gasket looks the same at every scale, the same tensor encodes each of its length scales, so the size of the representation grows only linearly with the fractal generation. Using this encoding, we study a Chern insulator, the flux kagome model, confined to this fractal geometry, and develop methods to investigate its topological states within a tensor network framework. This allows us to reach much higher fractal generations than would otherwise be possible with dense-matrix methods, resolving the density of states of a deep fractal with over 100 million sites. Furthermore, the larger structures allow us to resolve topology in smaller gaps, suggesting that these methods could be pivotal in understanding the interplay between topology and non-integer dimensionality. 
\end{abstract}
	
\maketitle{}

\section{Introduction} 
Topological phases of matter are ubiquitous in condensed matter physics \cite{thouless_quantized_1982,haldane_model_1988,hasan_colloquium_2010,qi_topological_2011,chiu_classification_2016}. They are often characterized in terms of momentum-space invariants and, for finite samples, edge states and topological invariants are related via the bulk-boundary correspondence \cite{hatsugai_chern_1993}. However, in many situations, a real-space description is necessary, because the spatial periodicity underpinning momentum space descriptions is inherently broken. This occurs,  for instance in amorphous \cite{Agarwala2017TopologicalSystems,Grushin2022TopologicalMatter} or aperiodic \cite{kraus_topological_2012,Verbin2013ObservationQuasicrystals,bandres_topological_2016,Zilberberg2021TopologyInvited,fan_topological_2022,moustaj_anomalous_2025,tiago_tensor_2026} systems. Fractal geometries, which constitute a specific example of such aperiodic structures, have in recent years garnered interest, as they test the limits of several of the assumptions underlying our understanding of these phases of matter.
Previous studies have shown that they can host, for instance, Chern insulating phases \cite{canyellas_topological_2024,osseweijer_haldane_2024,brzezinska_topology_2018,iliasov_hall_2020,fremling_existence_2020,fischer_robustness_2021,yang_photonic_2020,biesenthal_fractal_2022,li_fractal_2023,salib_topological_2024}, or higher-order topological phases \cite{manna_higher-order_2022}, which fractal geometries can even induce \cite{eek_fractality-induced_2025}.
However, the fractal structure raises fundamental questions about the applicability of bulk-boundary correspondence, since in a fractal the distinction between bulk and boundary is often unclear, as can be reflected by its noninteger Hausdorff dimension \cite{hausdorff_dimension_1918,mandelbrot_fractal_1986}. The many ways topological phases show up in fractal systems therefore raise the question of what role scale-invariance plays. Answering it requires describing a fractal's structure and properties over many generations, but the number of sites grows multiplicatively from one generation to the next, which makes studying scale-invariant systems an exponentially hard numerical problem. 

Meanwhile, tensor networks (TNs) have been widely used in quantum many-body physics for decades to tackle exponentially large Hilbert spaces, providing accurate solutions to paradigmatic problems such as the Hubbard and Heisenberg models \cite{White1992DensityGroups,Cirac2021MatrixTheorems,Banuls2023TensorMap,Orus2014AStates,Schollwock2011TheStates,Stoudenmire2012StudyingGroup,Fishman2022TheCalculations,Fishman2022CodebaseITensor,Chan2016MatrixAlgorithms,Orus2019TensorSystems}. Their success relies on the fact that the physically relevant corner of the Hilbert space typically has limited entanglement between subsystems, which carefully crafted algorithms exploit to compress the problem efficiently into a network of small, locally connected tensors.
More recently, they have also been used to tackle more general large linear-algebra problems, leading to speedups of several orders of magnitude in many problems \cite{Waintal2026WhoTechniques,Zhou2020WhatComputers,Shinaoka2023MultiscaleTrains,Erpenbeck2023TensorModels,NunezFernandez2022LearningTrains,Takahashi2025CompactnessPropagators,Jeannin2025ComprehensiveBlockade,Rohshap2025Two-particleEquations}, ranging from computational chemistry \cite{Jolly2025TensorizedChemistry} to machine learning \cite{Stoudenmire2016SupervisedNetworks,Dilip2022DataLearning,Han2018UnsupervisedStates,Huggins2019TowardsNetworks} and dynamical simulations \cite{Peddinti2024,Gourianov2025,Niedermeier2026SolvingRepresentation,boucomasGP,chenGP,2025arXiv250701149C}.
They have also been used for large tight-binding problems, where quantics tensor cross interpolation (QTCI) \cite{Oseledets2010TT-crossArrays,Oseledets2011Tensor-TrainDecomposition,Ritter2024QuanticsFunctions,Jeannin2024Cross-extrapolationRegime,NunezFernandez2025LearningLibraries,Ritter2022QuanticsTCI.jl,Ritter2022TensorCrossInterpolation.jl} efficiently compresses real-space functions across very large lattices, with a memory cost that scales only as $\mathcal{O}(\log N)$ with the number of sites $N$ \cite{Fumega2025CorrelatedAlgorithm,Sun2025Self-consistentSites,tiago_tensor_2026,Moustaj2026TensorSystems,Moustaj2026Tensor-networkSites,Sun2026Real-spaceNetworks,Antao2026TensorApplications}.
These developments have enabled, among others, the study of quasicrystalline mosaics of topological Chern states \cite{tiago_tensor_2026}, the self-consistent solution of interacting super-moir\'e systems beyond one billion lattice sites \cite{Sun2025Self-consistentSites}, the direct computation of spatially dependent, momentum-resolved spectral functions of large nonperiodic systems \cite{Moustaj2026TensorSystems}, and even the solution of two-particle Bethe--Salpeter Hamiltonians for excitons in super-moir\'e and quasicrystalline systems \cite{Moustaj2026Tensor-networkSites}.
However, the QTCI encoding fails to compress the purely quasiperiodic substitution rules that generate quasicrystals, since a finite set of such rules cannot be generated in the binary number system it employs.
A recent work showed that, beyond approximate schemes like QTCI, tight-binding Hamiltonians of one-dimensional quasicrystals can be encoded as exact matrix product operators (MPOs), by leveraging their self-similar structure through finite-state automata \cite{Kolehmainen2026One-dimensionalAutomata}. 

Building on this previous work, we show that fractals are especially well suited for tensor network methods, precisely because of their scale-invariance. We encode each length scale of the lattice in one core of an MPO. Because a self-similar fractal looks the same at every scale, all cores but the last are then identical, and in scale space the fractal resembles a crystal.
We formulate an encoding that captures the scale-invariance exactly, and use it to study a nearest-neighbor tight-binding model on the Sierpi\'nski gasket, in which time-reversal symmetry is broken by Peierls phases.
Since the resulting MPO has the same small bond dimension at every scale, we can compute the density of states of a 17th-generation gasket with over 100 million sites. Projecting the spectral function onto the plane waves of the periodic kagome lattice reveals band-like and Dirac-like features, despite the absence of translational invariance. We then evaluate the real-space Chern number $\nu$ as an MPO contraction, combined with a hybrid method of purification and sparse diagonalization that is able to resolve narrow gaps, and find plateaus with $\nu=+1$ and $\nu=-1$. Finally, the local Chern marker and the local density of states, resolved scale by scale, show that the same parts of the lattice can behave as bulk in one gap and as edge in another, with edge-like structures appearing on different length scales.

During the writing of this manuscript, we became aware of a recent preprint that considers a similar idea, namely encoding self-similar systems efficiently as tensor networks \cite{Brzezinska2026Finite-stateLattices}.

\section{Tensor-network representation of the fractal Chern insulator}
\begin{figure}[!t]
    \centering
    \includegraphics[width=\linewidth]{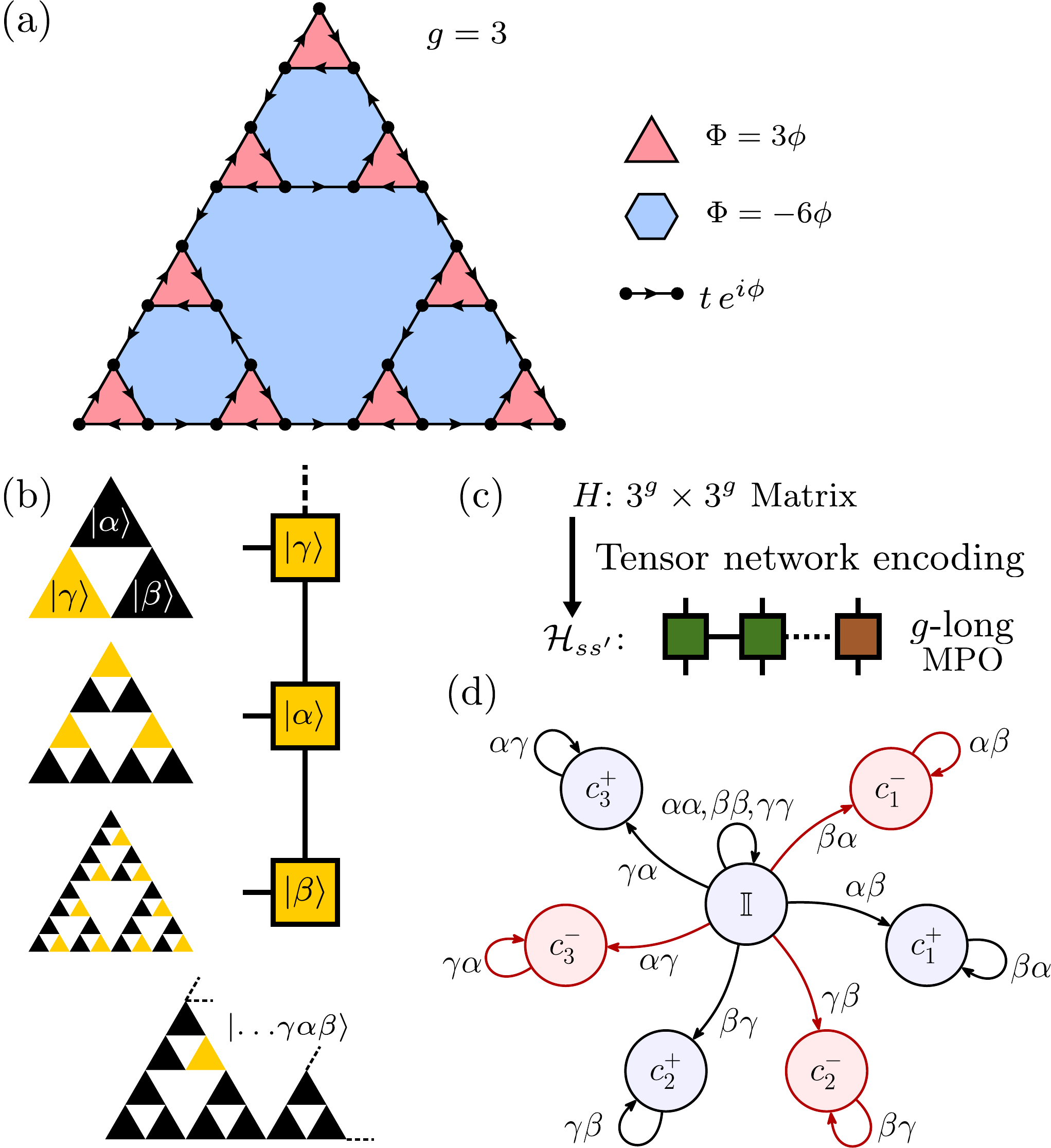}
    \caption{Using scale-invariance to separate the Hamiltonian into different length scales. (a) The flux kagome model of Eq.~\eqref{eq:H} on the $g=3$ SG, with its $27$ sites and nearest-neighbor bonds. A hop along an arrow carries the amplitude $t\,e^{i\phi}$, so every smallest triangle encloses the flux $3\phi$ (pink) and every hole the flux $-6\phi$ (blue), both counted clockwise. (b) The scale-separated encoding of a site. A site is selected by a sequence of choices, one per length scale, represented by an index sequence. (c) In this scale-separated formalism, the Hamiltonian of the $g$th generation SG is exactly representable as an MPO, with $g-1$ repeated geometry tensor cores (green) and one final model-specific (brown) tensor core. (d) Because the SG is self-similar, all geometry tensor cores are identical and can be represented as an automaton. Here, each circle represents a virtual state, and the arrows indicate what transition occurs for a given set of inputs. A missing arrow means that no hopping occurs, which gives a zero matrix element.}
    \label{fig:encoding}
\end{figure}

We consider a flux kagome model confined to the geometry of the Sierpi\'nski gasket (SG), a fractal obtained by recursively subdividing a triangle into four subtriangles and removing the central one. After $g$ iterations, one is left with the $g$th-generation SG. The Hamiltonian for this model reads 
\begin{equation}
    H = t\sum_{\langle i,j \rangle} e^{i\phi_{ij}}c^\dagger_ic_j ,
    \label{eq:H}
\end{equation}
where $t$ is the nearest-neighbor hopping amplitude, $\phi_{ij}$ is the corresponding Peierls phase, and $\langle i,j\rangle$ is the set of nearest-neighbor pairs on the fractal. The lattice and its fluxes are shown in Fig.~\ref{fig:encoding}(a). In two dimensions, the flux kagome model is a Chern insulator and hosts robust topological edge modes \cite{Ohgushi2000}. Previous work on topological systems on fractal lattices has revealed that the self-similar geometry and non-integer dimensionality can give rise to rich topological phase diagrams \cite{osseweijer_haldane_2024}. Here, we investigate the topological properties of the flux-kagome model on the Sierpi\'nski gasket, using a scale-invariant MPO encoding to push the system size beyond what was previously achievable.

An MPO is a one-dimensional chain of tensor cores \cite{Schollwock2011TheStates,Orus2014AStates,Chan2016MatrixAlgorithms}. When the site index is written digit by digit, each core acts on one digit and therefore on one length scale of the lattice \cite{Shinaoka2023MultiscaleTrains,Ritter2024QuanticsFunctions}. We propose to exploit this scale separation to encode fractal lattices, whose geometry is self-similar across length scales.
Figure \ref{fig:encoding}(b) depicts how the recursive structure of the SG provides a natural way to describe the geometry at successive scales. 
At each level, a three-valued index selects the top ($\alpha$), bottom-right ($\beta$), or bottom-left ($\gamma$) subtriangle. Repeating this choice at every scale results in an index sequence that specifies a single location on the gasket, from which its two-dimensional coordinates can be reconstructed. Using this scale-based encoding the Hamiltonian can be represented as an MPO.

Throughout this work, ordinary symbols denote physical operators and their matrix representations, while calligraphic symbols with a hat denote their corresponding MPO representations. Because of scale invariance, the MPO of the Hamiltonian in Eq.~\eqref{eq:H} is built from two types of tensor cores, as shown in Fig.~\ref{fig:encoding}(c). The first $g-1$ cores are identical geometry tensor cores (green). Reading the index sequences of two sites scale by scale, they record whether the two sites are nearest neighbors and along which edge of a triangle their bond lies. Each geometry tensor core encodes one length scale of the fractal, and since the gasket is self-similar, the same rule connects sites at every scale. The scale invariance of the gasket therefore becomes a translational invariance of the MPO. Moving one core along the chain corresponds to moving one length scale up or down in the fractal, and a larger generation only adds further identical cores. The final core (brown) reads the last pair of indices, which places the two sites within their smallest triangles, and attaches the hopping amplitude and Peierls phase of the bond, $t\,e^{\pm i\phi}$ depending on its type and direction. All parameters of Eq.~\eqref{eq:H} therefore sit in the final core.
This construction provides an exact and compact representation of the Hamiltonian, which reads

\begin{equation}
    \hat{\mathcal{H}} = \sum_{j\in\{1,2,3\}} \sum_{l=1}^{g}
    \left[ \bigotimes_{i=1}^{l-1} I \right]
    \sigma_j
    \left[ \bigotimes_{i=l+1}^{g} \sigma_j^\dagger \right] t_l
    + \mathrm{H.c.},
\end{equation}
\begin{figure*}
    \centering
    \includegraphics[width=\linewidth]{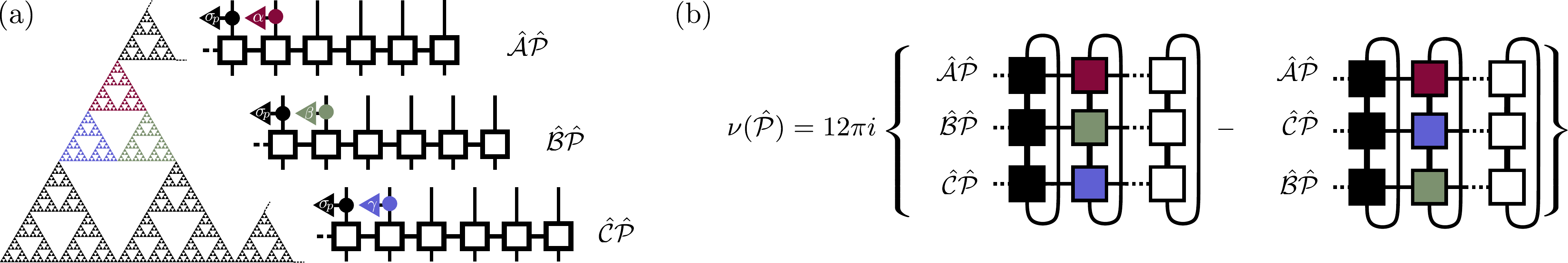}
    \caption{The real-space Chern number (RSCN) adapted to MPOs. (a) The subsets $A$ (red), $B$ (green) and $C$ (blue) and the construction of the corresponding subset projectors. (b) The MPO contraction resulting in the RSCN of Eq.~\eqref{eq:RSCN_Kitaev}.  }
    \label{fig:topoMarkers}
\end{figure*}
where $I$ is the identity operator on the three-dimensional local space spanned by $\{\alpha,\beta,\gamma\}$.
The operator $\sigma_j$ is the $j$th step of the clockwise cycle $\alpha\rightarrow\beta\rightarrow\gamma\rightarrow\alpha$, that is, $\sigma_1=\ket{\beta}\bra{\alpha}$, $\sigma_2=\ket{\gamma}\bra{\beta}$ and $\sigma_3=\ket{\alpha}\bra{\gamma}$, and $\sigma_j^\dagger$ is the reverse step.

The form of $\hat{\mathcal{H}}$ follows from how a bond appears in the index sequences of its two sites. Let $l$ be the first scale on which the two sequences differ. On the scales before $l$, the two sites lie in the same subtriangle, so their indices coincide and the MPO acts with $I$. On scale $l$, they lie in two different subtriangles of the same triangle, for example $\alpha$ and $\beta$, which $\sigma_1$ connects. On every finer scale, the bottom-right corner of the top triangle lies in its bottom-right subtriangle and the top corner of the bottom-right triangle in its top subtriangle, so the two sites are $\ket{\dots\alpha\beta\beta\cdots\beta}$ and $\ket{\dots\beta\alpha\alpha\cdots\alpha}$. After scale $l$, the pair of indices is therefore that of scale $l$ in reversed order, which is the step $\sigma_j^\dagger$ on every remaining core. The same holds for the other two pairs of subtriangles and, because the SG is self-similar, on every scale $l$, so the same few operators generate all bonds between subtriangles. A bond inside a smallest triangle is the case $l=g$, in which the two sites differ only on the last scale and the product of $\sigma_j^\dagger$ is empty. The final core of Fig.~\ref{fig:encoding}(c) is therefore the factor on the last scale, $\sigma_j$ if $l=g$ and $\sigma_j^\dagger$ if $l<g$, multiplied by $t_l$, the hopping amplitude and Peierls phase of the bond. For the flux kagome model of Eq.~\eqref{eq:H}, $t_g=t\,e^{i\phi}$ and $t_l=t\,e^{-i\phi}$ for $l<g$. The Hermitian conjugate adds the reverse hoppings.
This construction is general, since the geometry tensor cores encode only the connectivity of the SG. Other nearest-neighbor models on the gasket, with different hopping amplitudes and Peierls phases for each type of bond, are therefore obtained by changing the final core alone. More generally, any self-similar lattice described by a finite-state automaton can be encoded in the same way \cite{Kolehmainen2026One-dimensionalAutomata,Brzezinska2026Finite-stateLattices}.


The set of geometry tensor cores can also be represented as an automaton \cite{Crosswhite2008FiniteAlgorithms,Kolehmainen2026One-dimensionalAutomata,Brzezinska2026Finite-stateLattices}, as shown in Fig.~\ref{fig:encoding}(d). Here, $\bra{i}H\ket{j}$ can be evaluated by reading the index sequences of $i$ and $j$ scale by scale. One starts in the initial virtual state $\mathbb{I}$. At each scale, the pair of indices determines which transition to follow. After such a transition, the automaton is in one of seven virtual states, namely the initial state $\mathbb{I}$ or one of six connectivity states, $c_i^+$ (clockwise) and $c_i^-$ (counterclockwise). The index $i$ labels the edge along which the connection lies,
with $i=1$ for the top-right, $i=2$ for the bottom, and $i=3$ for the top-left edge. If no transition exists for a combination, the corresponding matrix element vanishes. After all geometry tensor cores have been traversed, the resulting virtual state, $\mathbb{I}$ for a bond inside a smallest triangle and a connectivity state $c_j^\pm$ for a bond between subtriangles, forms the input for the final tensor.
The small number of virtual states sets the bond dimension of $\hat{\mathcal{H}}$. It is at most seven for every generation $g$, so the memory cost of $\hat{\mathcal{H}}$ grows only linearly with $g$, that is, logarithmically with the number of sites $N=3^g$.

\section{Topological markers}
Quantifying the topological nature of fractals cannot be done using the usual momentum-space invariants because of their inherent lack of translational invariance in real space. Instead, topology must be characterized using real-space methods. Here, we employ two complementary approaches. The real-space Chern number (RSCN) \cite{kitaev_anyons_2006} provides a global measure of topology, and the local Chern marker \cite{bianco_mapping_2011} resolves the local topological profile across the lattice.

The RSCN has been shown to work well in aperiodic systems \cite{Chen2023KitaevInsulators} and in particular in fractal geometries \cite{osseweijer_haldane_2024}. 
It is given by
\begin{equation}
    \nu({P}) = 12\pi i \sum_{j \in A}\sum_{k \in B}\sum_{l \in C} \left(P_{jk}P_{kl}P_{lj} - P_{jl}P_{lk}P_{kj}\right),
    \label{eq:RSCN_Kitaev}
\end{equation}
where $P_{ij}$ is a matrix element of the Fermi sea projector, and $j$, $k$, and $l$ are site indices restricted to three touching subsets $A$, $B$ and $C$ of the lattice arranged in clockwise order. We denote the Fermi energy by $E$ and write $P(E)$ for the corresponding Fermi projector, which projects onto all eigenstates with energy below $E$. In the tensor-network formalism, $P(E)=\theta(E-H)$, with $\theta$ the Heaviside step function, is obtained directly as an MPO through a purification algorithm, which starts from a linear function of $\hat{\mathcal{H}}$ and iterates a polynomial map until the result is idempotent \cite{McWeeny1960SomeTheory,Niklasson2002ExpansionMatrix}. Refs.~\cite{tiago_tensor_2026,Antao2026TensorApplications} provide the details of the algorithm. In this work, we use a slightly modified approach described in Sec.~\ref{sec:spectral}.

Here, we introduce a tensor network approach to calculating the RSCN, in which each subset is represented with scale-separated indices. Specifically, selecting appropriate subsets reduces to fixing the first tensor cores to a common prefix of $p$ indices, followed by an $\alpha$, $\beta$ or $\gamma$ index for the $A$, $B$ and $C$ subsets, respectively. The remaining $l$ indices are left free, so that each subset forms an $l$th-generation Sierpi\'nski gasket, as shown in Fig.~\ref{fig:topoMarkers}(a). The corresponding subset projectors are constructed by contracting $\hat{\mathcal{P}}$ with the appropriate Kronecker-delta tensors (circles) and identity tensors (straight lines). In a regime where the RSCN is quantized, its value should be insensitive to the precise placement of the subsets \cite{osseweijer_haldane_2024}. 

Using the subset projectors, the RSCN can be evaluated as an MPO contraction,
\begin{equation}
\nu(\hat{\mathcal{P}}) = 12\pi i \Tr{
\hat{\mathcal{A}}\hat{\mathcal{P}}\hat{\mathcal{B}}\hat{\mathcal{P}}\hat{\mathcal{C}}\hat{\mathcal{P}} 
-  \hat{\mathcal{A}}\hat{\mathcal{P}}\hat{\mathcal{C}}\hat{\mathcal{P}}\hat{\mathcal{B}}\hat{\mathcal{P}}
},    
\end{equation}
which is depicted diagrammatically in Fig.~\ref{fig:topoMarkers}(b). Here, $\hat{\mathcal{A}}$, $\hat{\mathcal{B}}$ and $\hat{\mathcal{C}}$ are the MPOs of the projectors onto the three subsets, for example $\sum_{j\in A}\ket{j}\bra{j}$ for $A$. Each is a product over the cores, with a projector onto the fixed index on each of the first $p+1$ cores and the identity on the remaining $l$ cores, drawn as the Kronecker-delta tensors and lines in Fig.~\ref{fig:topoMarkers}(a). The contraction above is therefore equivalent to Eq.~\eqref{eq:RSCN_Kitaev}, but written with MPOs.

While the RSCN provides a global measure of topology, the local Chern marker resolves the spatial distribution of the topological response. It has been studied in aperiodic systems \cite{Tran2015TopologicalQuasicrystal,tiago_tensor_2026} and applied to fractal geometries \cite{Lage2025TopologicalCarpet,Lage2026OrbitalFractals}. Ref.~\cite{tiago_tensor_2026} demonstrates how a quenched local Chern marker represented as an MPO can be used to spatially resolve topological phases in two-dimensional systems. It is defined as
\begin{equation}
    \mathcal{C}_i = 2\pi i\bra{i}
    \hat{\mathcal{Q}} \hat{\mathcal{X}}_i^\Lambda \hat{\mathcal{P}} \hat{\mathcal{Y}}_i^\Lambda \hat{\mathcal{Q}}
    -
    \hat{\mathcal{P}} \hat{\mathcal{X}}_i^\Lambda \hat{\mathcal{Q}} \hat{\mathcal{Y}}_i^\Lambda \hat{\mathcal{P}}
    \ket{i},
\end{equation}
where $\hat{\mathcal{P}}$ and $\hat{\mathcal{Q}}=\mathds{1}-\hat{\mathcal{P}}$ are MPO representations of the projectors onto occupied and unoccupied states, respectively, and $\hat{\mathcal{X}}_i^\Lambda$ and $\hat{\mathcal{Y}}_i^\Lambda$ are the MPO representations of the quenched position operators $X_i^\Lambda=\Lambda\sin[(X-x_i)/\Lambda]$ and $Y_i^\Lambda=\Lambda\sin[(Y-y_i)/\Lambda]$. Here, $X$ and $Y$ are the position operators, $(x_i,y_i)$ is the position of site $i$, and $\Lambda$ is a length chosen large compared with the correlation length of $P$ \cite{tiago_tensor_2026}. In this work, we set $\Lambda=5$ in units of the gasket's side length. Close to site $i$, $X_i^\Lambda\approx X-x_i$ and $Y_i^\Lambda\approx Y-y_i$, so $\mathcal{C}_i$ reduces to the marker of Ref.~\cite{bianco_mapping_2011}, while the bounded operators keep the bond dimension of their products with $\hat{\mathcal{P}}$ small; see Ref.~\cite{tiago_tensor_2026} for details. 

Although the local Chern marker has been studied in periodic \cite{bianco_mapping_2011,Bau2024LocalSystems} and aperiodic two-dimensional systems \cite{Tran2015TopologicalQuasicrystal,tiago_tensor_2026}, as well as on bulky fractal geometries \cite{Lage2025TopologicalCarpet,Lage2026OrbitalFractals}, its interpretation requires care. In finite systems, the marker can exhibit substantial boundary contributions. On a fractal lattice, the intricate geometry and extensive boundary can make it difficult to distinguish bulk topological response from boundary effects. We therefore use the RSCN to identify topological regimes and subsequently evaluate the local Chern marker within these regimes to investigate the spatial distribution of the topological profile.

\section{The spectral properties and global topology}\label{sec:spectral}
\begin{figure*}
    \centering
    \includegraphics[width=\linewidth]{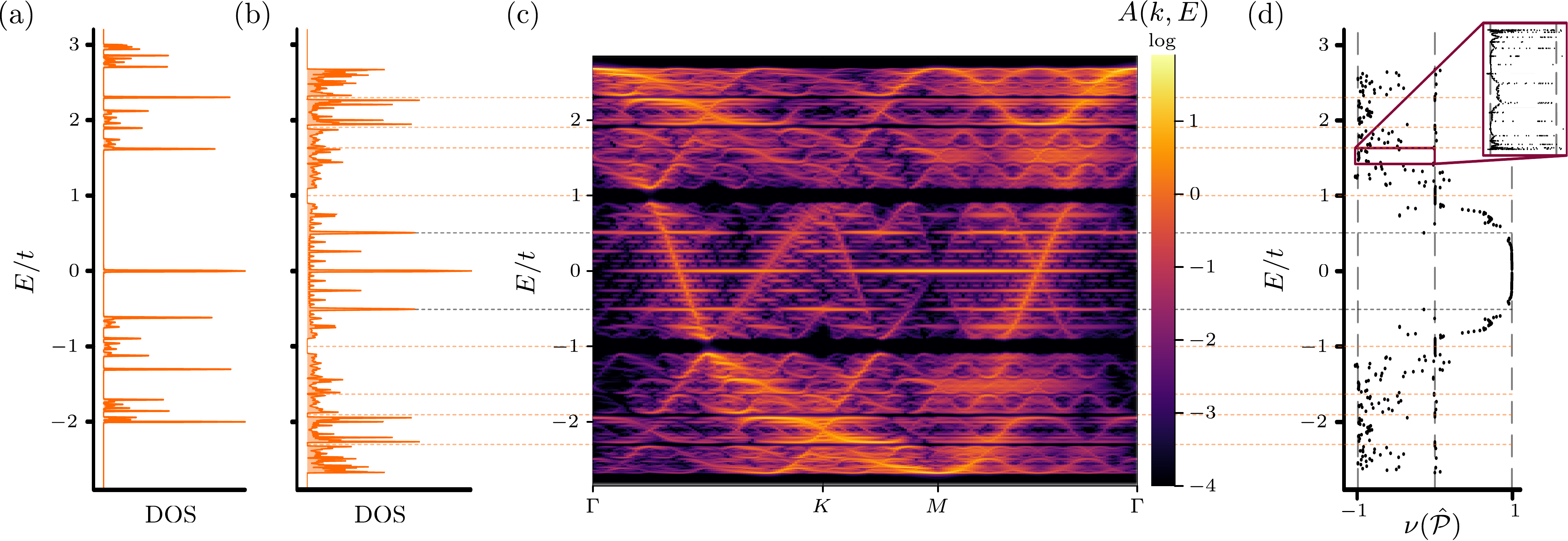}
    \caption{Density of states (DOS) of the Sierpi\'nski kagome model, computed for a 17th-generation gasket. (a) Zero flux, $\phi=0$. (b) Finite flux, $\phi=\pi/6$. (c) The momentum-resolved spectral function $A(k,E)$ of the Sierpi\'nski kagome model on a 10th-generation gasket, obtained by projecting the spectral function, computed with the kernel polynomial method (KPM), onto plane-wave states along the high-symmetry path $\Gamma-K-M-\Gamma$ of the periodic kagome lattice. The spectral weight reveals band-like features despite the absence of translational invariance. (d) RSCN of the same 10th-generation gasket, evaluated for the Fermi projector $\hat{\mathcal{P}}(E)$ as a function of the Fermi energy $E$. The inset shows a zoom of the RSCN obtained using a hybrid method combining MPO-based purification with sparse exact diagonalization on a larger region.}
    \label{fig:DOS}
\end{figure*}
At zero flux, the model reduces to a nearest-neighbor kagome model. The density of states (DOS) of a 17th-generation SG with over 100 million sites, computed with the kernel polynomial method (KPM) \cite{Weisse2006TheMethod}, is shown in Fig.~\ref{fig:DOS}(a). Unlike the translationally invariant two-dimensional model, the fractal system exhibits large spectral gaps driven by the fractal geometry of the lattice. When a non-zero flux is introduced, the spectrum spreads out, the gaps become narrower, and a more intricate, self-similar structure becomes apparent in the DOS, as depicted in Fig.~\ref{fig:DOS}(b).

A strikingly similar pattern has been observed for other topological models on the SG \cite{osseweijer_haldane_2024,li_fractality-induced_2023}, consistent with the spectrum being dominated by the geometry tensor cores, which all these models share, rather than by the model-specific final core. 
In particular, the initially flat band at $E=0$ spreads across the energy interval $E\in[-1,1]$ when a finite flux is introduced. Within this interval, a self-similar pattern of high-DOS (bulk-like) and low-DOS (in-gap-like) regions emerges.

To further investigate how the fractal geometry modifies the spectrum, we project the spectral function onto the Brillouin zone of the periodic kagome lattice. 
Although translational invariance is absent and momentum is not a good quantum number, we can still use plane waves as a projection basis.  A similar approach was shown to spatially resolve band-like structures in two-dimensional nonperiodic systems encoded as tensor networks \cite{Moustaj2026TensorSystems}. 
This is particularly natural here, as the underlying lattice retains the local structure and point-group symmetry of the kagome lattice, while the fractal geometry emerges on length scales larger than the nearest-neighbor bond length. We therefore expect the spectral features associated with both the underlying kagome structure and the fractality to be captured within its first largest hexagon that would correspond to Brillouin zone of the periodic lattice.

Figure~\ref{fig:DOS}(c) shows the momentum-resolved spectral function of the Sierpi\'nski kagome model along the high-symmetry path $\Gamma-K-M-\Gamma$ of the kagome lattice. It shows a remarkably rich structure. First, we see that the gaps marked by orange dashed lines in Fig.~\ref{fig:DOS}(b) correspond to true gaps with no spectral weight along this high-symmetry path. Additionally, in the range $E\in [ -1, 1 ]$ we see extended flat signals at the energies corresponding to the flat bands of the DOS in Fig.~\ref{fig:DOS}(b). 
On top of those flat features, diagonal structures resembling linearly dispersive modes are observed. 

Throughout the plot, a recurrent pattern of crossing linear spectral weight appears, reminiscent of Dirac-like features. The largest of these are found in the central region, although the different branches have different weights. At energies further from zero, the Dirac-like features become smaller, but more of them appear, evocative of the self-similar patterns in the geometry. 

Because these low-DOS regions form as a direct consequence of the flux term, which drives topology in two dimensions, and seem to host dispersive spectral weight, they are candidate gaps in which topological phases may occur. 
To establish their topological character, the RSCN of the Fermi projector is evaluated as a function of energy in Fig.~\ref{fig:DOS}(d). 

\begin{figure*}[t]
    \centering
    \includegraphics[width=\linewidth]{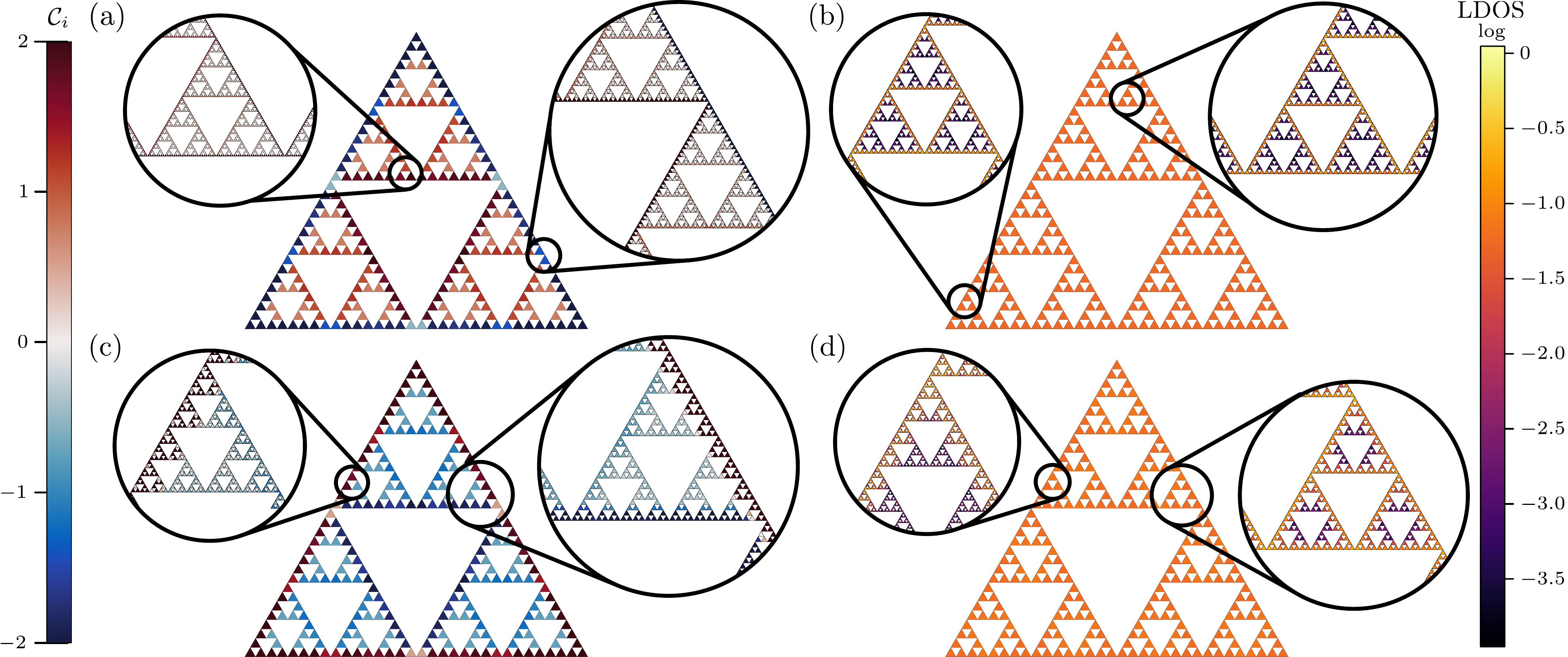}
    \caption{The local Chern marker and LDOS of a 10th-generation gasket, evaluated on different scales at two different energies, $E=-0.178$ and $E=1.567$. The plotted resolution is controlled by the number $s$ of final tensor cores contracted together. Each displayed element therefore represents a region of the underlying lattice, with smaller $s$ corresponding to finer resolution. The insets show smaller subregions selected by fixing a longer prefix with smaller $s$, allowing finer structures to be resolved. In the main panels, $s=5$, so each element represents a fifth-generation subfractal. (a) The local Chern marker $\mathcal{C}_i$ evaluated at $E=-0.178$, where $\nu(\hat{\mathcal{P}})=+1$. Here, both insets show atomic resolution ($s=0$). (b) The logarithm of the LDOS corresponding to the same energy considered in (a). Again, the insets show the atomically resolved LDOS ($s=0$), showing a small subset of the lattice. (c) $\mathcal{C}_i$ evaluated at $E=1.567$, where $\nu(\hat{\mathcal{P}})=-1$. Here, the left inset shows atomic resolution ($s=0$), and the right inset has $s=1$, to resolve the larger structures. (d) The logarithm of the LDOS in the $E=1.567$ gap. The lattice is similarly resolved with $s=0$ and $s=1$ in the left and right insets, respectively.  }
    \label{fig:localC}
\end{figure*}

Obtaining a well-converged Fermi projector is the main computational bottleneck of this methodology. 
The self-similar spectrum contains narrow gaps between bulk-like bands, while edge-like states can populate these gaps when the system is in a topological regime. Resolving these gaps with KPM-based approaches \cite{Weisse2006TheMethod} is computationally demanding at the energy resolution these gaps require. McWeeny \cite{McWeeny1960SomeTheory} and SP2 \cite{Niklasson2002ExpansionMatrix} purification can also converge slowly or fail to converge when the spectral separation is insufficient, as is often the case when a gap is populated with topological modes. Consequently, obtaining RSCN curves at high energy resolution was challenging even for a 10th-generation SG.

In contrast, the DOS vanishes in relatively large regions, indicating an enhanced spectral separation, so that purification converges reliably within the spectral gaps marked by orange dashed lines. Therefore, we developed a hybrid method that first obtains the Fermi projector at one of these reference energies using SP2 purification in the tensor-network representation. To reach the target filling or energy, the missing eigenstates between the reference energy and the target are then obtained using sparse exact diagonalization and incorporated into the projector.
This approach requires storing the sparse Hamiltonian, sacrificing some of the memory advantage of the tensor-network method. However, only a limited number of additional eigenstates (on the order of $10^3$) need to be computed, making this approach substantially faster than solving directly for all occupied eigenstates. The hybrid method thus combines the tensor-network representation of the reference projector with sparse diagonalization to achieve energy resolution on par with exact diagonalization, without the computational cost of calculating all eigenvectors. 

The main panel of Fig.~\ref{fig:DOS}(d) uses SP2 purification alone, but the inset evaluates the RSCN using larger regions and utilizes the hybrid method to achieve the required resolution. Here, the RSCN exhibits a rich energy dependence, indicative of the self-similar gap structure. In particular we focus on a broad plateau of $\nu(\hat{\mathcal{P}})=+1$ on either side of the flat band at $E=0$ and a narrow plateau of $\nu(\hat{\mathcal{P}})=-1$ around $E\approx1.5$, shown in the inset. The plateau around the flat band is interrupted by sharp drops of the RSCN coinciding with sharp peaks in the DOS (indicated by black lines), suggestive of their bulk-like character. Similar behavior is visible in the higher-resolution inset around $E\approx1.5$.

Note that the region used to calculate the RSCN in the inset was much larger than in the main panel, i.e., the prefix in Fig.~\ref{fig:topoMarkers}(a) was much shorter.

\section{Spatially resolved topology}
To investigate whether this difference in quantization behavior is related to the spatial structure of the states, we examine the local Chern marker and local density of states (LDOS) at energies corresponding to the two stable RSCN plateaus. 
Figure~\ref{fig:localC} shows the local Chern marker and LDOS at these two energies. The LDOS is obtained from the eigenvectors calculated using the sparse eigensolver in the hybrid method. Together, these quantities provide complementary information. The local Chern marker resolves the spatial distribution of the topological response, while the LDOS reveals where the states have spectral weight.
The spatial resolution is controlled by the tensor-network representation. Because each core encodes one length scale, averaging the LDOS to a given length scale $s$ amounts to contracting the last $s$ cores. In the main panels, the final five tensor cores are contracted, yielding fifth-generation images in which each displayed triangle represents the average over the last five length scales. The insets show subsets of the fractal at different resolutions, controlled by the number $s$ of contracted final tensor cores.

Figures \ref{fig:localC}(a) and (b) display the local Chern marker and LDOS at $E=-0.178$ with RSCN $\nu(\hat{\mathcal{P}}) \approx 0.998$. In the main panel of Fig.~\ref{fig:localC}(a), the interior exhibits extended regions of near-quantized values of $\mathcal{C}_i \approx 1$ (red). In contrast, the outer boundary shows a large negative contribution $\mathcal{C}_i < -1$ (blue).  Furthermore, an interior edge with a large opposite sign contribution, $\mathcal{C}_i > 1$ (dark red), is also visible. The insets show a subset of the lattice at atomic resolution ($s=0$), i.e. each triangle represents a single site. Here, the local Chern marker concentrates on the interior boundaries, while the sites surrounding progressively smaller holes carry a smaller local Chern marker, $0<\mathcal{C}_i<1$.
In Fig.~\ref{fig:localC}(b) the logarithmic LDOS at the same energy is depicted. The main panel shows barely any spatial variation because each of the fifth-generation subtriangles contains comparable amounts of low- and high- LDOS regions. This becomes apparent in the insets, where the atomic resolution ($s=0$) reveals a strong enhancement of the LDOS around fourth generation fractals. 
The higher-energy gap at $E=1.567$, where $\nu(\hat{\mathcal{P}})\approx-0.989$, is investigated in Figs.~\ref{fig:localC}(c) and (d). The main panel resembles a sign-inverted version of Fig.~\ref{fig:localC}(a). A large interior region exhibits $\mathcal{C}_i\approx-1$, while the interior and outer boundaries carry enhanced contributions of opposite sign. However, the insets reveal an important distinction. Whereas the opposite-sign contributions in Fig.~\ref{fig:localC}(a) are concentrated on single lines of sites, here they extend over much larger structures. The left inset of Fig.~\ref{fig:localC}(c) shows the local Chern marker at atomic resolution ($s=0$), while the right inset shows it on one length scale larger ($s=1$). In the left inset an entire third-generation fractal exhibits enhanced, edge-like local Chern marker values, rather than the single line of sites observed in Fig.~\ref{fig:localC}(a). At $s=1$, the opposite-sign contributions along the interior and outer boundaries appear on comparable length scales. Although the signal seems to decay over a second-generation SG, each displayed triangle at this resolution already represents an average over first-generation fractals.
The LDOS in Fig.~\ref{fig:localC}(d) also shows enhancement over larger structures. At $s=0$, the left inset reveals enhanced spectral weight across an entire third-generation SG, while the right inset ($s=1$) shows that the LDOS is concentrated around the boundaries of fifth-generation subfractals.  This supports the interpretation that the higher-energy gap exhibits edge-like topology on a larger length scale than in the lower-energy gap.

We hypothesize that these larger edge-like structures of both the local Chern marker and the LDOS explain why a larger region was required to obtain a well-quantized RSCN in the higher energy gap. Because the subsets $A$, $B$ and $C$ involved in calculating the RSCN are necessarily situated near boundaries at some scale, capturing the topological nature of these longer-length-scale features requires larger regions.
We further conjecture that this behavior is related to the scale invariance of the fractal geometry. At different energies, similar edge-like structures could recur across different length scales; in the infinite-generation limit of the fractal, their absence at a particular scale would appear to break this self-similarity.

Although a clean distinction between bulk and boundary contributions is difficult to make in a fractal geometry, the local Chern marker and LDOS reveal that different parts of the lattice can exhibit bulk- or edge-like behavior depending on the energy. In particular, the edge-like structures forming in the high-energy gap consist of atoms that showed more bulk-like behavior in the low-energy gap. Additionally, the boundaries at the holes of the 5th generation, which exhibit edge-like behavior in the low-energy gap, appear bulk-like in the high-energy gap. 

\section{Conclusion} 
We have developed a tensor-network formalism for self-similar tight-binding models, in which each core of the MPO represents one length scale, and demonstrated it with an exact and compact encoding of Hamiltonians on the Sierpi\'nski gasket. Identical geometry tensor cores with seven virtual states encode the self-similar geometry and a single final core encodes the model details, so the bond dimension does not grow with the generation. The geometric self-similarity of the lattice is thus reflected directly in the translational invariance of the tensor cores along the scale direction. Viewed in scale space, the fractal is effectively a crystal. For the flux kagome model, this allowed us to compute the DOS of a 17th-generation gasket with over 100 million sites, which shows large fractal gaps and a self-similar structure. We also projected its spectral function onto the plane waves of the periodic kagome lattice, where band-like features persist despite the absence of translational invariance. The momentum-space representation also shows intricate self-similar Dirac-like structures at various energies, which may correspond to self-similar edge states. Evaluating the RSCN as an MPO contraction, together with a hybrid of SP2 purification and sparse diagonalization, we identified plateaus with $\nu=+1$ and $\nu=-1$. The local Chern marker and the LDOS, resolved scale by scale, showed that whether a part of the lattice behaves as bulk or as edge depends on the energy, and that the edge-like structures of the $\nu=-1$ gap extend over larger length scales, consistent with the larger regions needed to quantize the RSCN there.

Because the geometry and the model enter through separate cores, changing the hopping amplitudes or phases only requires a different final core, and other self-similar lattices can be encoded through their finite-state automata \cite{Kolehmainen2026One-dimensionalAutomata,Brzezinska2026Finite-stateLattices}. The encoding is also compatible with existing tensor-network methods for large tight-binding problems, which makes interactions at the mean-field \cite{Sun2025Self-consistentSites} or two-particle level \cite{Moustaj2026Tensor-networkSites} on fractals accessible. The main limitation is the convergence of the Fermi projector in narrow gaps, which currently requires the hybrid scheme used here. Lifting it would allow following the scale-dependent edge-like structures to deeper generations and testing whether they recur across length scales, as the self-similarity of the gasket suggests. More broadly, the self-similarity that makes fractals exponentially costly for conventional numerics is precisely the structure that makes them natural for tensor networks, suggesting a promising route toward characterizing topology in scale space, much as momentum space does for translationally invariant systems.

\section{Acknowledgements} 
We acknowledge the computational resources provided by the Aalto Science-IT project and the financial support from  InstituteQ,
the Research Council of Finland (project No. 370912 and the Finnish Quantum Flagship project No. 358877, Aalto University), the Finnish Ministry
of Education and Culture through the Quantum Doctoral Education Pilot Program (QDOC VN/3137/2024-OKM-4),
the Finnish Centre of Excellence in Quantum Materials QMAT (No. 374166),
and the ERC Consolidator Grant ULTRATWISTROICS (Grant agreement no. 101170477). 
The authors thank M. Kolehmainen for insightful discussions on the automaton representation of MPSs and MPOs. We also thank L. Eek for useful discussions on the interpretation of topology in fractals. The code used for this work can be consulted at~\cite{FractalMPOrepo}. We used Claude Code (Anthropic), with the Opus 5.5 and Fable 5.2 models, for code debugging, optimization, and proofreading.

\bibliography{bibliography,bib_anouar}

\end{document}